\documentclass[sigconf, screen]{acmart}
\renewcommand\footnotetextcopyrightpermission[1]{} % removes footnote with conference information in first column
\pdfoutput=1%For ARXIV
\usepackage{algorithm}
\usepackage{algorithmic}
\usepackage{multirow}
\usepackage{subfigure}
\usepackage{pifont}
\usepackage{array}
\usepackage{seqsplit}
\newcolumntype{C}[1]{>{\centering\arraybackslash}p{#1}}

\AtBeginDocument{%
  }

\setcopyright{acmlicensed}
\copyrightyear{2025}
\acmYear{2025}
\acmDOI{XXXXXXX.XXXXXXX}
\acmConference[MM '25]{Make sure to enter the correct
  conference title from your rights confirmation email}{October 27-- October 31,
  2025}{Dublin, Ireland}
\acmISBN{978-1-4503-XXXX-X/2025/06}

\begin{document}
\title{Lightweight Medical Image Restoration via Integrating Reliable Lesion-Semantic Driven Prior}

\author{Pengcheng Zheng}
\affiliation{%
  \institution{University of Electronic Science and Technology of China}
  \city{Chengdu}
  \country{China}
}
\email{zpc777@std.uestc.edu.cn}

\author{Kecheng Chen}
\affiliation{%
  \institution{City University of Hong Kong}
  \city{Hong Kong}
  \country{China}
}
\email{cs.ckc96@gmail.com}
\authornote{Corresponding authors.\label{cor-author}}

\author{Jiaxin Huang}
\affiliation{%
  \institution{Mohamed bin Zayed University of Artificial Intelligence}
  \city{Abu Dhabi}
  \country{The United Arab Emirates}
}
\email{hjx1134@gmail.com}

\author{Bohao Chen}
\affiliation{%
  \institution{University of Electronic Science and Technology of China}
  \city{Chengdu}
  \country{China}
}
\email{202322080702@std.uestc.edu.cn}

\author{Ju Liu}
\affiliation{%
  \institution{University of Electronic Science and Technology of China}
  \city{Chengdu}
  \country{China}
}
\email{2019270103004@std.uestc.edu.cn}

\author{Yazhou Ren}
\affiliation{%
  \institution{University of Electronic Science and Technology of China}
  \city{Chengdu}
  \country{China}
}
\affiliation{%
  \institution{Shenzhen Institute For Advanced Study, University of Electronic Science and Technology of China}
  \city{Shenzhen}
  \country{China}
}
\email{yazhou.ren@uestc.edu.cn}

\author{Xiaorong Pu\textsuperscript{\ref {cor-author}}}
\affiliation{%
  \institution{University of Electronic Science and Technology of China}
  \city{Chengdu}
  \country{China}
}
\affiliation{%
  \institution{Shenzhen Institute For Advanced Study, University of Electronic Science and Technology of China}
  \city{Shenzhen}
  \country{China}
}
\email{puxiaor@uestc.edu.cn}
% \authornote{Corresponding authors.\label{cor-author}}
% \author{Yu Zhu}
% \email{yuzhu@nwpu.edu.cn}
% \affiliation{%
%   \institution{Northwestern Polytechnical University}
%   \city{Xi'an}
%   \country{China}
% }
% \authornote{Corresponding author}

% \author{Qingsen Yan}
% \email{qingsenyan@nwpu.edu.cn}
% \affiliation{%
%   \institution{Northwestern Polytechnical University}
%   \city{Xi'an}
%   \country{China}
% }
% \author{Jinqiu Sun}
% \email{sunjinqiu@nwpu.edu.cn}
% \affiliation{%
%   \institution{Northwestern Polytechnical University}
%   \city{Xi'an}
%   \country{China}
% }
% \author{Yanning Zhang}
% \email{ynzhang@nwpu.edu.cn}
% \affiliation{%
%   \institution{Northwestern Polytechnical University}
%   \city{Xi'an}
%   \country{China}
% }
% %
% % By default, the full list of authors will be used in the page
% % headers. Often, this list is too long, and will overlap
% % other information printed in the page headers. This command allows
% % the author to define a more concise list
% % of authors' names for this purpose.

% %
% % The abstract is a short summary of the work to be presented in the
% article.
\begin{abstract}
Medical image restoration tasks aim to recover high-quality images from degraded observations, exhibiting emergent desires in many clinical scenarios, such as low-dose CT image denoising, MRI super-resolution, and MRI artifact removal. Despite the success achieved by existing deep learning-based restoration methods with sophisticated modules, they struggle with rendering computationally-efficient reconstruction results. Moreover, they usually ignore the reliability of the restoration results, which is much more urgent in medical systems. To alleviate these issues, we present \textbf{LRformer}, a \textbf{L}ightweight Trans\textbf{former}-based method via \textbf{R}eliability-guided learning in the frequency domain. Specifically, inspired by the uncertainty quantification in Bayesian neural networks (BNNs), we develop a \textbf{R}eliable \textbf{L}esion-Semantic \textbf{P}rior \textbf{P}roducer (\textbf{RLPP}). RLPP leverages Monte Carlo (MC) estimators with stochastic sampling operations to generate sufficiently-reliable priors by performing multiple inferences on the foundational medical image segmentation model, MedSAM. Additionally, instead of directly incorporating the priors in the spatial domain, we decompose the cross-attention (CA) mechanism into real symmetric and imaginary anti-symmetric parts via fast Fourier transform (FFT), resulting in the design of the \textbf{G}uided \textbf{F}requency \textbf{C}ross-\textbf{A}ttention (\textbf{GFCA}) solver. By leveraging the conjugated symmetric property of FFT, GFCA reduces the computational complexity of naive CA by nearly half. Extensive experimental results in various tasks demonstrate the superiority of the proposed LRformer in both effectiveness and efficiency.
\end{abstract}





\maketitle

\newcommand{\eg}{\textit{e}.\textit{g}. }
\section{Introduction}
Medical images can provide crucial information to improve diagnostic effectiveness, such as identifying lesions, monitoring diseases, and guiding treatments \cite{brown2005distributed, gholipour2010robust, schnabel2003validation}. However, due to inherent imaging mechanism, medical images often suffer from various forms of degradation during acquisition, significantly impacting diagnostic accuracy. Thus, medical image restoration (MedIR) task plays an indispensable role in recovering the high-quality (HQ) image from its degraded low-quality (LQ) counterpart. 
\begin{figure}[!t]
\centering
\hspace{-3.7mm}
\includegraphics[width=3.4in]{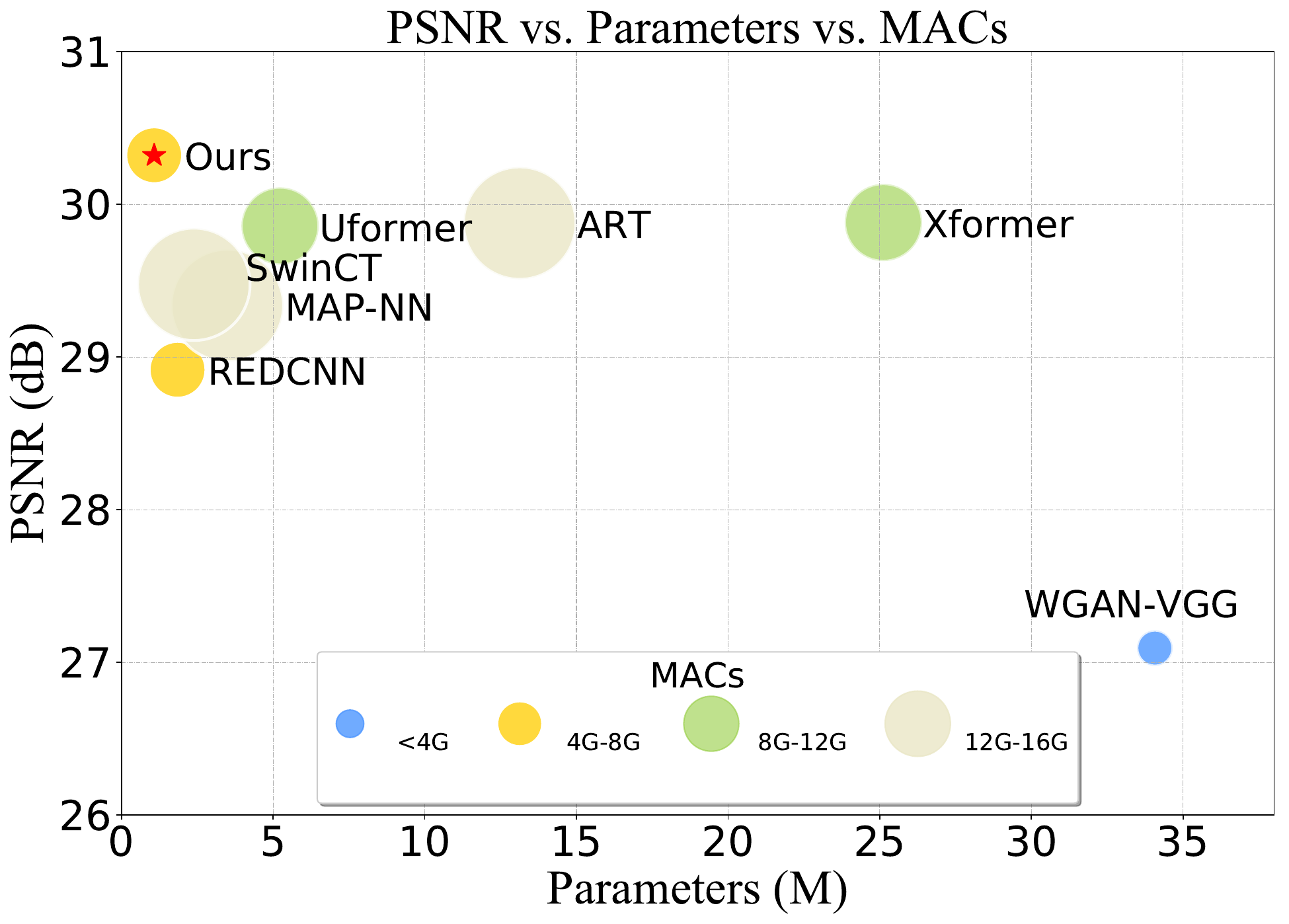}
\caption{PSNR results v.s the Parameters and MACs of different methods for LDCT image denoising on the AAPM dataset, demonstrating the superior performance of LRformer.}
\label{F1}
\end{figure}

In the era of deep learning, convolution neural networks (CNNs) have been extensively investigated for image restoration. Numerous CNNs-based methods achieve promising reconstruction results via multi-stage \cite{MAPNN, RCAN, FBCNN, BlockCNN}, multi-scale \cite{EDSR} and adversarial learning \cite{WGAN-VGG} schemes. However, as a local information descriptor, CNNs struggle with their restricted receptive field, failing to capture long-range pixel dependencies \cite{sattler2019understanding}. To overcome this limitation, the Transformer model \cite{vaswani2017attention} has emerged as a powerful alternative for image restoration, utilizing the self-attention (SA) mechanism to learn high-order features for enhanced contextual representation \cite{high-order}. A variety of works, such as Uformer \cite{Uformer}, ART \cite{ART}, SwinIR \cite{SwinIR}, and Restormer \cite{Restormer}, leverage Transformers for image restoration, leading to state-of-the-art (SOTA) performance.

While existing image restoration methods have proven effective for natural images, they suffer from a heavily-computational cost of scaled dot-product attention in Transformer models, which increases quadratically with image spatial resolution. Especially, this heavily-computational cost is especially terrible when dealing with medical images. For example, the CT imaging for one patient contains hundreds of high-resolution slices, making these methods infeasible to implement in resource-restricted scenarios. Although several existing MedIR methods attempt to improve the efficiency of the naive Transformer \cite{Hformer, guo2023reconformer}, they struggle to provide useful priors to guide the image restoration process, resulting in undesirable performance. A potential solution may derive from the lesion-semantic driven priors (\textit{e.g.}, lesion detection and semantic segmentation results) given by the downstream tasks, thereby providing useful guidance to enhance the restoration quality \cite{chen2021lesion, hu2022unified}. However, such a strategy still ignores the reliability of these priors due to the structural information loss of degraded images, leading to the trustworthiness risk of medical systems. To conclude, it is still challenging for existing MedIR methods to render a computationally-efficient reconstruction results in the context of sufficient reliability.

To address these limitations, we focus on efficiently restoring degraded medical images in the context of sufficient reliability. To this end, we propose to leverage a probabilistic framework endowed by variational Bayesian inference to acquire more reliable priors. By exploiting the powerful generalizability of MedSAM \cite{MesSAM} across various types of medical images, we first design a reliable lesion-semantic prior producer (RLPP), which performs multiple inferences of the probabilistic MedSAM model to yield sufficiently-reliable priors, which can provide lesion-semantic and shape information to improve the restoration performance. Although existing spatial domain-based attention mechanism can reduce computational complexity of naive Transformer, such as window-based \cite{SwinIR} and linear Transformer \cite{MaxSR}, the former lacks global information, and the latter cannot capture abundant 2D representations. Therefore, to achieve a computationally-efficient guidance process for the MedIR task, we propose a novel guided frequency cross-attention (GFCA) solver to embed the reliable priors in the frequency domain. Finally, we integrate these approaches into an end-to-end learnable framework, coined as LRformer. As depicted in Figure \ref{F1}, the proposed LRformer not only requires fewer model parameters and multiply-accumulate operations (MACs), but also achieves a superior restoration performance, compared with SOTA counterparts.

The main contributions of this paper are as follows:
\begin{enumerate}
\item[$\bullet$] By employing Monte Carlo (MC) estimators in the MedSAM model, a reliable lesion-semantic prior producer (RLPP) is proposed to directly generate sufficiently-reliable priors from degraded medical images. The segmentation results can provide lesion-semantic and shape priors to guide the restoration process. 
\item[$\bullet$] 
To efficiently integrate the acquired reliable priors into the image restoration process, we propose a novel guided frequency cross-attention (GFCA) solver, which can significantly reduce nearly half computational complexity of the naive CA by leveraging the conjugated symmetric property of FFT.
\item[$\bullet$] 
By combining reliable priors with the efficient GFCA solver, we develop the LRformer, a lightweight and reliable medical image restoration framework via reliability-guided learning in the frequency domain. More importantly, we provide a theoretical analysis to prove its superiority in computational efficiency.
\item[$\bullet$]
Extensive experimental results demonstrate that our proposed LRformer achieves SOTA performance in various MedIR tasks, such as LDCT denoising, MRI super-resolution, and MRI artifact removal.
\end{enumerate}

\begin{figure*}[!t]
\begin{minipage}[b]{1.0\linewidth}
  \centering
  \centerline{\includegraphics[width=17.8cm]{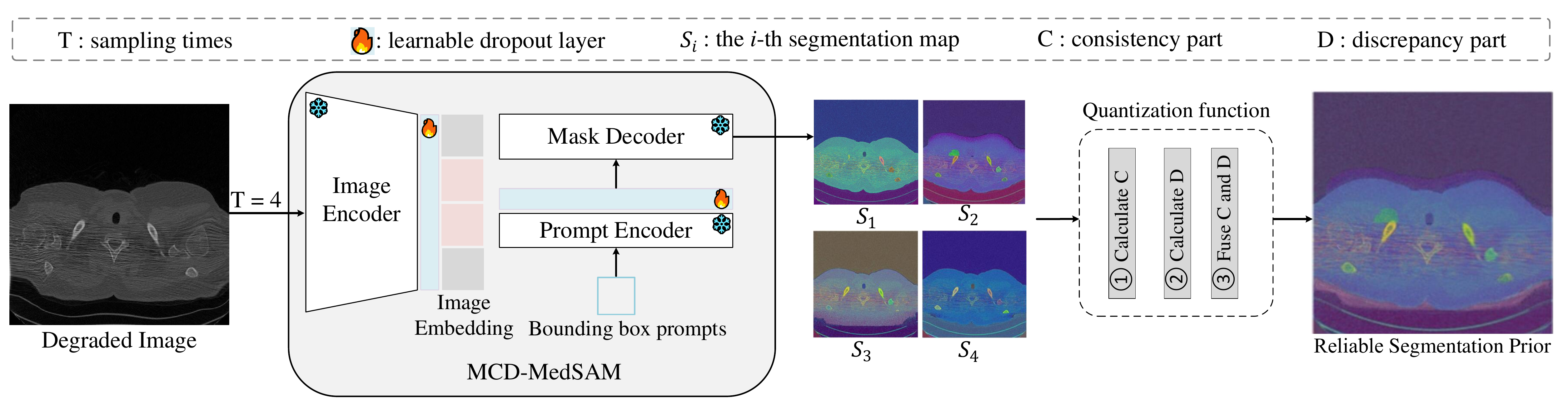}}
\end{minipage}
\caption{The computing flow of the reliable lesion-semantic prior producer based on MedSAM. We use the bounding box prompts with the shape of $[H, W]$ to segment the whole image.}
\label{USPP}
\end{figure*}

\vspace{-0.3cm}
\section{Related works}

\subsection{Uncertainty-Guided Learning}
Recently, advances \cite{uncertainty1, Uuncertainty2, uncertainty3} have been made in exploiting uncertainty to guide model learning and improve performance. Uncertainty is useful in machine learning. It can identify where a model is likely to make incorrect predictions and when input data is out-of-distribution (OOD) \cite{clyde2004model, onatski2003modeling}. A traditional method for estimating uncertainty is the BNNs, which attempt to learn a distribution over each of the model’s weight parameters. However, Bayesian inference is computationally intractable for these models in practice. Therefore, recent works often exploit MC estimators to produce uncertainty estimates. For example, Chen et al. \cite{chen2024unsupervised} propose leveraging MC samples to quantify model uncertainty and improve image reconstruction performance through uncertainty alignment. Evan Hann et al. \cite{hann2021ensemble} present a new framework utilizing MC sampling to provide robust segmentation results with inherent quality control.

% Different from these methods, in this work, we are the first attempt to generate reliable priors for the MedIR task via uncertainty quantification. 
In contrast to existing methods, we are the first to generate reliable priors for the MedIR task through uncertainty quantification.

\subsection{Image Restoration with Transformer}
% As the Transformer \cite{vaswani2017attention} model can model long-range contextual representations, it has been widely developed to tackle the image restoration task. 
Transformer \cite{vaswani2017attention}, known for its ability to model long-range contextual relationships, has been widely adopted for image restoration tasks. To improve the efficiency of the naive Transformer, several mutations \cite{MaxSR, MaxVIT, Restormer, SwinTransformer} are proposed. There are generally two types of Transformer as window-based and linear Transformer. Window-based Transformer performs the SA mechanism within a restricted window to capture the local bias. For example, SwinIR \cite{SwinIR} employs SA in an 8 $\times$ 8 local window for feature extraction and achieves excellent performance in natural image restoration tasks. As to the linear Transformer, Yang et al. \cite{MaxSR} propose to use adaptive MaxViT block \cite{MaxVIT} for image super-resolution, which decomposes the 2D SA into 1D, thereby reducing the complexity of SA to $o(n)$.
% As to the linear Transformer, Yang et al. \cite{MaxSR} propose to use adaptive MaxViT block \cite{MaxVIT} for image super-resolution, which decomposes the 2D SA into 1D and therefore reduces the complexity of SA into $o(n)$. 

Although these methods have demonstrated promising performance, they have limitations for MedIR tasks. The window-based Transformer often lacks global information, while the linear Transformer cannot capture abundant 2D representations. To address these shortcomings, we develop an efficient guided frequency cross-attention solver, which explores the conjugated symmetric property of FFT to efficiently integrate the acquired priors.

\section{Methodology}

\begin{figure*}[!t]
\begin{minipage}[b]{1.0\linewidth}
  \centering
  \centerline{\includegraphics[width=18cm]{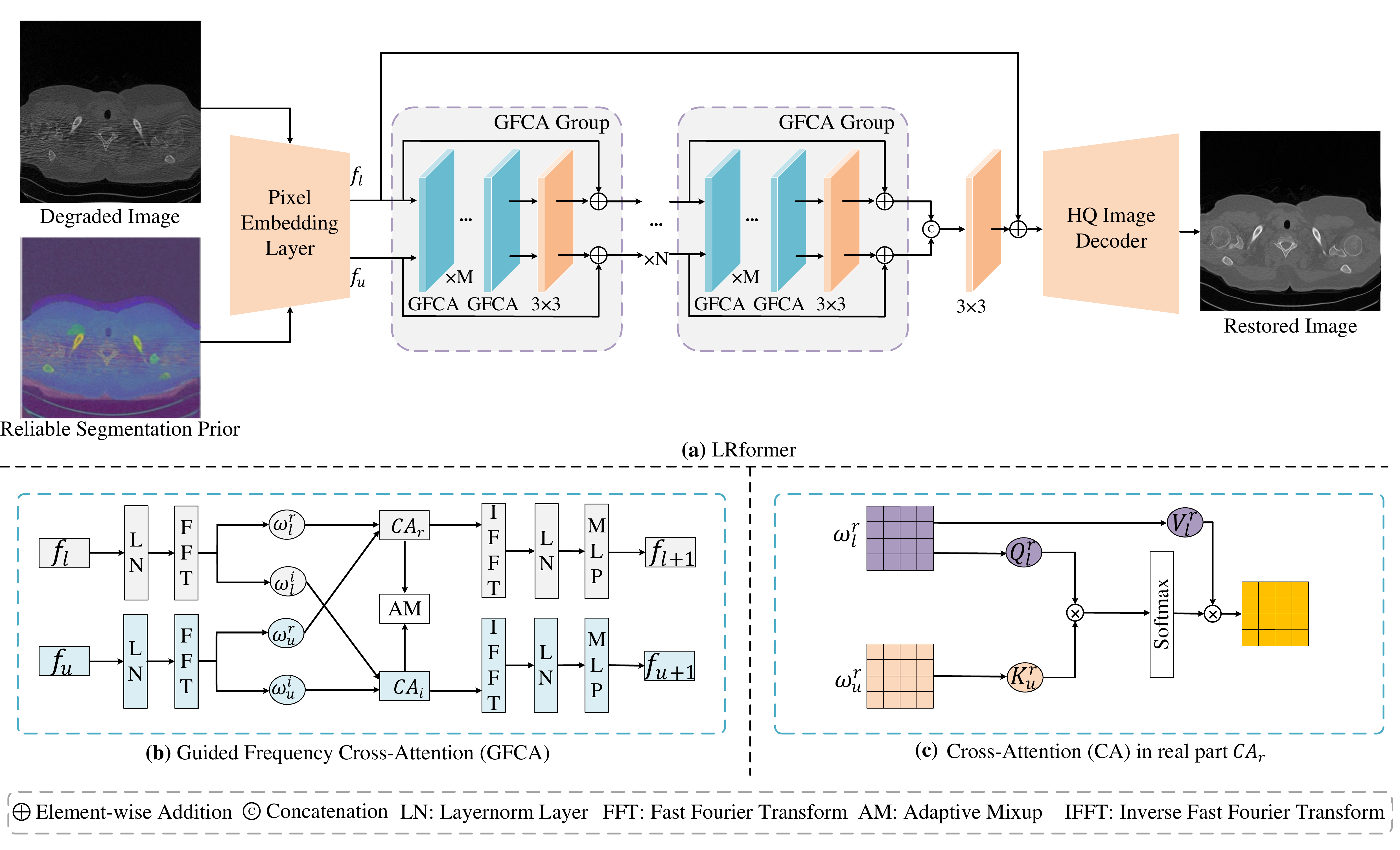}}
\end{minipage}
\caption{Top part (a) is the overview architecture of the proposed LRformer. Bottom-left (b) is the diagram of the guided frequency cross-attention (GFCA) solver. Bottom-right (c) takes the calculation process of the designed cross-attention (CA) mechanism in real part $CA_r$ as an example, and the diagram of $CA_i$ is as same as $CA_r$.}
\label{EUFformer}
\end{figure*}

We aim to present a reliable and lightweight method for addressing the MedIR task. By leveraging the explicit lesion-semantic and shape priors of segmentation maps, MedIR methods tend to achieve better reconstruction results. To this end, we first design a reliable lesion-semantic prior producer (RLPP) to directly generate sufficiently-reliable priors from degraded medical images. Furthermore, to efficiently exploit the priors estimated by the RLPP, we develop a guided frequency cross-attention (GFCA) solver, which explores the conjugated symmetric property of FFT to efficiently integrate the priors via the novel CA mechanism. Finally, we formulate these approaches into an end-to-end trainable network to address the MedIR task. The overview architecture of the proposed LRformer is shown in Figure \ref{EUFformer}(a). The details of each component are described as follows.

\subsection{Reliable Lesion-Semantic Prior Producer}
Existing MedIR methods struggle to provide sufficiently-reliable priors from downstream tasks. Therefore, we design the reliable lesion-semantic prior producer (RLPP) to directly generate reliable priors from degraded LQ images. Especially, the notable advantage of directly segmenting LQ degraded images, as opposed to clean images, lies in the fact that this approach does not require ground-truth data. This characteristic aligns well with realistic medical clinical scenarios, where it is often not feasible to scan a patient more than once using imaging equipment. By leveraging the powerful generalizability of the MedSAM model, RLPP can provide sufficiently-reliable priors for various types of medical images.

Although the MedSAM model is effective, its predicted results are deterministic, which decreases reliability in clinical scenarios, especially for segmenting degraded LQ medical images. We thus design a BNNs-based MedSAM model, which employs distributional modeling of the network weights to capture the segmentation uncertainty. This purpose can be achieved using MC estimators with $T$ stochastic samples: 
\begin{equation}
    \mathbb{E}_{q\left(\mathbf{y}^* \mid \mathbf{x}^*\right)}\left(\mathbf{y}^*\right) \approx \frac{1}{T} \sum_{t=1}^T \widehat{\mathbf{y}}^*\left(\mathbf{x}^*, \mathbf{W}_1^t, \ldots, \mathbf{W}_L^t\right),
\end{equation}
where $\mathbf{W}_l^t$ denotes $t$-th Dropout sampling of the $l$-th neural network parameters. $\mathbf{x}^*$, $\mathbf{y}^*$ and $\widehat{\mathbf{y}}^*$ represent the input of the model, the target label and the predicted output, respectively. $q\left(\mathbf{y}^* \mid \mathbf{x}^*\right)$ represents the conditional probability distribution of the output $\mathbf{y}^*$ given the input $\mathbf{x}^*$. Since the vanilla MedSAM does not utilize a Dropout layer, we append two extra learnable Dropout layers into the MedSAM model where its original weights are frozen, creating the MCD-MedSAM, with weights $\mathbf{w} \sim P(\mathbf{w})$ (as shown in Figure \ref{USPP}). By performing multiple ($T=4$) MC samplings, we obtain different segmentation maps, represented as $S_{i}\;(i={1,2, ..., T})$. For the ``unconfident" parts of the input degraded image, MCD-MedSAM lacks sufficient epistemic capacities, resulting in more varied segmentation results. We assume that these ``unconfident" parts require more attention during image restoration as their semantic information is hard to understand. Therefore, we focus not only on the consistent parts of the segmentation results but also on the discrepancy parts. Consequently, we design the Quantization function to acquire the final reliable segmentation prior from $T$ segmentation results. The calculation process can be formulated as:
\begin{equation}
\begin{gathered}
C=\sum_\cup\left(S_i \cap S_j\right), \quad
D=\sum_\cup\left(\left|S_i-S_j\right|\right), 
\end{gathered}
\end{equation}
\begin{equation}
\begin{gathered}
U=\alpha * C+\beta * D, 
\end{gathered}
\end{equation}
where $S_i$ is the $i$-th segmentation map. $C$ and $D$ are the consistent and discrepant parts of these segmentation maps, respectively, $i, j \in\{1,2, \cdots, T\}$. $\alpha$ (set as 0.5), and $\beta$ (set as 0.5) are weighted coefficients, representing the contributions of each part.

\subsection{Guided Frequency Cross-Attention Solver}
The computational complexity of the naive CA mechanism grows quadratically with increasing image resolution, making it difficult to implement in commercial medical services. Thus, it is important to develop an efficient CA mechanism to integrate the sufficiently-reliable priors. Existing efficient attention mechanisms are primarily categorized into two paradigms: window-based attention and linear attention. Although these methods can directly reduce computational complexity in the spatial domain, window-based attention often leads to global information deficiency, while linear attention fails to capture abundant two-dimensional contextual representations. Therefore, we propose an innovative CA mechanism to efficiently integrate the acquired priors in the frequency domain. Concretely, our approach stems from the conjugated symmetric property of FFT:
\begin{equation}
\label{Eq4}
X[m]=\overline{X[N-m]},
\end{equation}
where $x[n]$ is a real-valued signal with $N$ samples. In other words, Eq. \ref{Eq4} indicates that a feature in the real spatial domain with the shape $[H, W]$ can be mapped into the complex frequency domain with the compressed shape $[H, W/2]$ without losing information. Therefore, we develop the guided frequency cross-attention solver (GFCA) to efficiently integrate the reliable lesion-semantic priors in the frequency domain instead of the spatial domain. This simple design can save three-quarters of the computational burden and it can be formulated as: 
\begin{equation}
\label{Eq3}
O\left(\left(HW\right)^2\right) \rightarrow O\left(1/4\left(HW\right)^2\right),
\end{equation}

As shown in Figure \ref{EUFformer}(b), the input degraded image feature $f_l\in\mathbb{R}^{H\times W\times C}$ and reliable lesion-semantic prior feature $f_u\in\mathbb{R}^{H\times W\times C}$ are first processed by the LayerNorm layer \cite{layernorm}. Then, their size is flattened into $[C\times HW]$ and transformed into the frequency domain via FFT. Next, we separate the complex values $\omega_l$ and $\omega_u$ into their real symmetrical part $\omega_l^r$, $\omega_u^r\in \mathbb{R}^{C\times(HW/2)}$ and imaginary anti-symmetrical part $\omega_l^i$, $\omega_u^i\in \mathbb{R}^{C\times(HW/2)}$. Subsequently, the dual-branch cross-attention is calculated on both real part and imaginary part as follows:
\begin{equation}  \operatorname{Attention}_r=\operatorname{softmax}\left(Q\left(\omega_l^r\right) K\left(\omega_u^r\right)^T\right) V\left(\omega_l^r\right),
\end{equation}
\begin{equation}
\operatorname{Attention}_i=\operatorname{softmax}\left(Q\left(\omega_l^i\right) K\left(\omega_u^i\right)^T\right) V\left(\omega_l^i\right),
\end{equation}
The attention map computation involves the matrix multiplication of $Q\left(\omega_l^s\right) K\left(\omega_u^s\right)^T, s \in\{r, i\}$, and the attention scores $\operatorname{Attention}_r$ and 
 $\operatorname{Attention}_i \in \mathbb{R}^{(H W / 2) \times(H W / 2)}$. Therefore, the computational complexity of the proposed CA is only $O\left(1/4\left(HW\right)^2\right)$. 
 
 It is worth noting that the dual-branch CA is calculated independently for the real and imaginary parts, leading to a lack of contextual interaction. To address this, we design the adaptive mixup (AM) operation to exchange information flow adaptively between the real symmetric part and the imaginary anti-symmetric part, formulated as:
 \begin{equation}
\operatorname{CA}_r=\sigma(\theta)*\operatorname{Attention}_r+(1-\sigma(\theta))*\operatorname{Attention}_i,
\end{equation}
\begin{equation}
\operatorname{CA}_i=\sigma(\theta)*\operatorname{Attention}_i+(1-\sigma(\theta))*\operatorname{Attention}_r,
\end{equation}
where $\sigma(\theta)$ is a learnable factor to fuse features between the dual branches, with its value determined by applying the sigmoid operator $\sigma$ applied to parameter $\theta$. Finally, the extracted features are transformed back into the spatial domain via inverse fast Fourier transform (IFFT) and subsequently input into the LayerNorm layer \cite{layernorm} and Multilayer Perceptron (MLP) \cite{MLP} Layer.

\begin{figure}[!t]
\centering
\hspace{-3.7mm}
\includegraphics[width=3.4in]{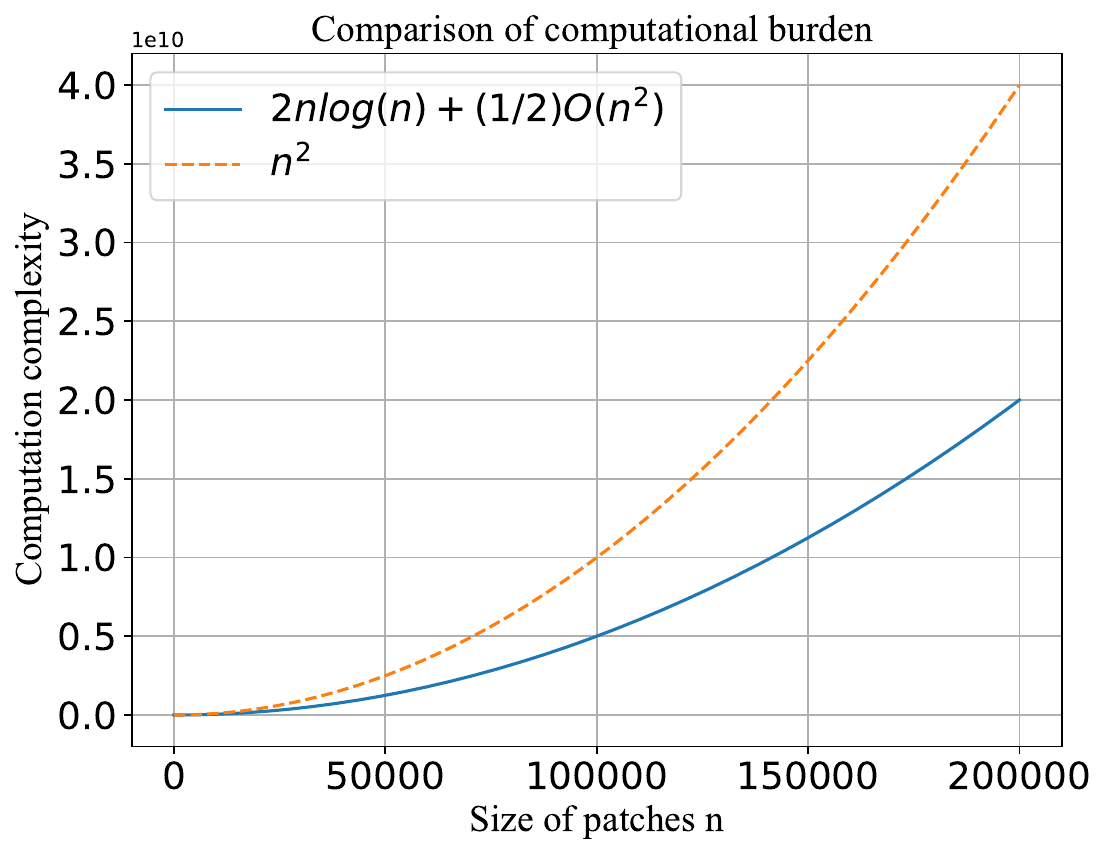}
\caption{Computational burden comparison of naive Transformer and the proposed LRformer. The computational superiority of LRformer over naive Transformer becomes more pronounced with the increasing of token-size $n$.}
\label{ComplexityFunction}
\vspace{-2.2em}
\end{figure}

\subsection{Lightweight Reliable Frequency Network}
The overall architecture of our proposed LRformer is shown in Figure \ref{EUFformer}(a), consisting of three parts: a pixel embedding layer, a deep feature encoder, and a high-quality (HQ) image decoder. Specifically, the pixel embedding layer employs a 3$\times$3 convolution layer to project the input images to shallow feature embeddings. Then, the shallow features undergo a hierarchical deep feature encoder to generate deep feature embeddings. It consists of $N$ guided frequency cross-attention (GFCA) groups, each containing $M$ GFCA blocks followed by a 3$\times$3 convolution with residual connections \cite{ResNet}. A 3$\times$3 convolution layer is added at the end of the deep feature encoder, providing a better foundation for the aggregation of shallow and deep feature embeddings. For the HQ image decoder, we use a simple convolution layer to reconstruct images for tasks that do not need upsampling, such as denoising and artifact removal. For the super-resolution task, a sub-pixel convolution layer \cite{shi2016real} is employed to upsample features. We optimize the parameters of LRformer by minimizing the $L_1$ pixel loss between the restored and ground-truth HQ images.

\subsection{Theoretical Analysis of Model Efficiency}
In this section, we provide a theoretical analysis of the computational efficiency of the proposed LRformer. Compared to the naive discrete Fourier transform (DFT), the computational complexity of FFT and its inverse process (IFFT) is only $O(nlog(n))$ \cite{FFTCom}. By leveraging the conjugated symmetric property of FFT, a tensor of shape $[C, n]$ can be mapped into the frequency domain with a compressed shape $[C, n/2]$. Therefore, the attention map calculated in GFCA has a reduced shape of $[n/2, n/2]$, where $n$ represents the patch size. The CA in both real symmetric and imaginary anti-symmetric parts has a computational cost of $(1/4) O(n^2)$ each. Thus, the total computational complexity of LRformer can be expressed as:
\begin{equation}
  2nlog(n) + 2\times(1/4)O(n^2)=2nlog(n)+(1/2)O(n^2),  
\end{equation}
which is more token-efficient than the vanilla CA with $O(n^2)$ computational complexity. Notably, this advantage becomes more significant as $n$ grows larger. A simple way to prove this is to visualize and compare their graphs of computational burden functions. As shown in Figure \ref{ComplexityFunction}, it can be obviously observed that the magnitude of deviation $\Delta n = n^2-[2nlog(n)+(1/2)O(n^2)]$ is gradually increasing with the increasing of token-size $n$, which strongly demonstrates that the excellent superiority of the proposed LRformer in terms of model efficiency.

\begin{table*}[!h]
    \small
    \renewcommand\arraystretch{1.1}
    \caption{The average quantitative results for various MedIR tasks. The best results are \textbf{bolded}, and the second-best results are \underline{underlined}. Note that the multiply-accumulate operations (MACs) of all models are tested on a degraded image with 64 $\times$ 64 pixels.}
    \begin{tabular*}{\textwidth}{@{\extracolsep{\fill}}cccccccc@{}}
    %\begin{tabular*}{\textwidth}{@{\extracolsep{\fill}} | l | c | %r |}
    \toprule
        Tasks & Methods & Params (M) & MACs (G) & PSNR ↑ & SSIM ↑ & LPIPS ↓ &  VIF ↑ \\ \hline
        &  MAP-NN \cite{MAPNN} & 3.49 & 13.79 & 29.3377 & 0.8529 & 0.0991 & 0.9204 \\
        &  REDCNN \cite{REDCNN} & 1.85 & 5.05 & 28.9162 & 0.8536 & 0.1247 & 0.8894 \\ 
        & WGAN-VGG \cite{WGAN-VGG} & 34.07 & 3.61 & 27.0933 & 0.8558 & 0.1383 & 0.8892 \\ 
        \multirow{2}{*}{LDCT Denoising} &SwinCT \cite{swinct} & 2.39 & 12.78 & 29.4759 & 0.8504 & 0.1119 & 0.9346 \\ 
        &Xformer \cite{Xformer} & 25.12 & 10.32 & \underline{29.8811} & \underline{0.8615} & 0.0952 & 0.9369 \\ 
        & Uformer \cite{Uformer} &5.23 & 12.00 & 29.8580 & 0.8608 & \underline{0.0902} & 0.9371 \\ 
        & ART \cite{ART} & 3.11 & 13.13 & 29.8768 & 0.8601 & 0.0986 & \underline{0.9391} \\ 
        & Ours & 1.31 & 7.96 & \textbf{30.3213} & \textbf{0.8663} & \textbf{0.0891} & \textbf{0.9407}
        \\ \midrule
        & Bicubic \cite{Bicubic} & \textemdash & \textemdash & 26.9123 & 0.8672  & 0.2591  & 0.9117 \\ 
        & EDSR \cite{EDSR} & 1.51 & 8.12 & 31.2690 & 0.9183 & 0.0860 & 0.9331 \\ 
        & RCAN \cite{RCAN} & 12.61 & 53.16 & 32.6248  & 0.9356  & 0.0814 & 0.9362\\ 
        \multirow{2}{*}{\quad MRI Super-Resolution} & SwinIR \cite{SwinIR} & 11.85 & 50.55 & 32.9600 & 0.9391 & 0.0760 & 0.9377\\ 
        & HAT \cite{HAT} & 20.51 & 86.4 & 32.9521 & 0.9391 & 0.0752 & 0.9374\\ 
        & Uformer \cite{Uformer} & 5.23 & 80.31  & \underline{32.9612} & 0.9387 & 0.0759 & 0.9379 \\ 
        & Restormer \cite{Restormer} & 26.09 & 88.18 & 32.9547 & \underline{0.9395} & \underline{0.0714} &\underline{0.9383}\\ 
        & Ours & 1.61 & 11.09 & \textbf{33.2298} & \textbf{0.9434} & \textbf{0.0652}  &\textbf{0.9395}\\ \midrule
            & WGAN-VGG \cite{WGAN-VGG} & 34.07 & 3.61 & 28.4915 & 0.8685  & 0.1064  & 0.8495 \\ 
        & FBCNN \cite{FBCNN} & 71.92 & 11.39 & 28.8449 & 0.8954 & 0.0625 & 0.8789 \\ 
        & BlockCNN \cite{BlockCNN} & 6.14 & 19.35 & 28.6021  & 0.8730  & 0.0988 & 0.8614\\ 
        \multirow{2}{*}{\quad MRI Artifact Removal} & Xformer \cite{Xformer}& 25.12 & 10.32 & 31.8866 & 0.9428 & 0.0448 & \underline{0.8829}\\ 
        & Uformer \cite{Uformer} & 5.23 & 12.00 & \underline{32.2862} & \underline{0.9479} & 0.0408 & 0.8623\\ 
        & Restormer \cite{Restormer} & 26.09 & 8.81  & 32.1044 & 0.9452 & 0.0414 & 0.8761 \\ 
        & ART \cite{ART}& 3.11 & 13.13 & 32.2607 & 0.9472 & \underline{0.0349} &0.8612\\ 
        & Ours & 1.31 & 7.96 & \textbf{32.5161} & \textbf{0.9490} & \textbf{0.0317}  &\textbf{0.8853}\\
        \bottomrule
    \end{tabular*}
    \label{SOTA compare}
\end{table*}

\begin{figure*}[!t]
\begin{minipage}[b]{1.0\linewidth}
  \centering
  \centerline{\includegraphics[width=18cm]{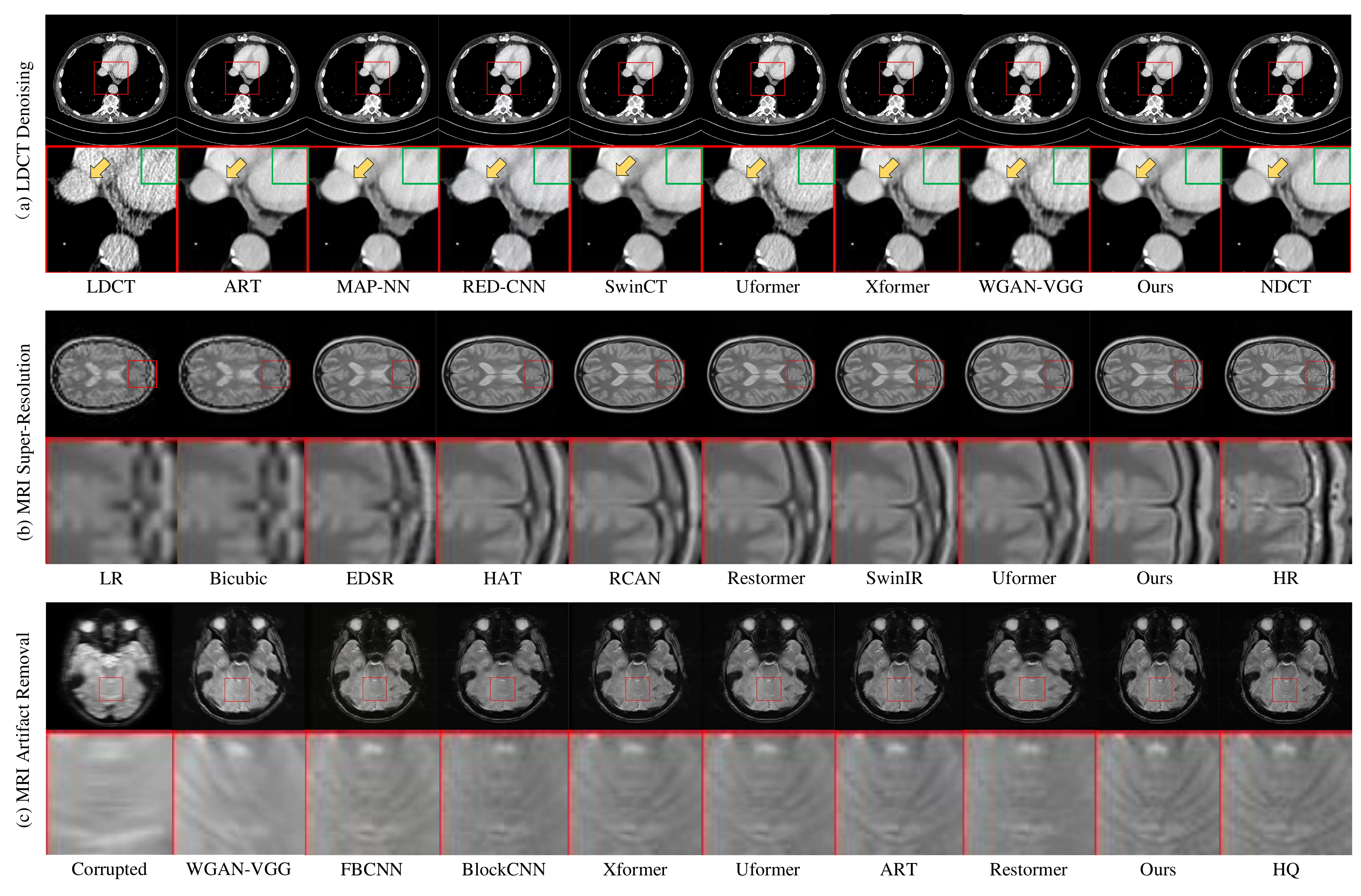}}
\end{minipage}
\caption{Visual comparison of different methods. (a) LDCT denoising. (The display window is [-160, 240] HU.) (b) MRI super-resolution. (c) MRI artifact removal. Zoomed ROI of the rectangle region is recommended for better visualization. Yellow arrows in (a) indicate regions with notable differences.}
\label{visual compare}
\end{figure*}

\section{Experimental Results}

\subsection{Datasets}
We conduct extensive experiments in various MedIR tasks to demonstrate the superiority of the proposed LRformer, including LDCT denoising, MRI super-resolution, and MRI artifact removal. The restoration performance is evaluated on ``2016 NIH-AAPM-Mayo Clinic Low-dose CT Grand Challenge” (AAPM-16) dataset \footnote{https://www.aapm.org/GrandChallenge/LowDoseCT/}, IXI-PD dataset \footnote{http://brain-development.org/ixi-dataset/} and ADNI dataset \footnote{https://adni.loni.usc.edu/data-samples/adni-data/}, respectively for these tasks.
(1) The AAPM-16 dataset serves as a widely adopted benchmark for LDCT denoising methods. It comprises 5,936 pairs of LDCT and NDCT images, each with a thickness of 1 mm, sourced from 10 patients. The LDCT images are simulated by introducing Poisson noise into the projection data prior to image reconstruction, with the noise level of the LDCT images corresponding to 25\% of that in the NDCT images. For our study, we selected 4,800 image pairs of 512 $\times$ 512 pixels from 8 patients for training, and 1,136 pairs from 2 additional patients for testing and validation. (2) The IXI-PD dataset comprises 500, 70, and 6 MR volumes designated for model training, testing, and quick validation, respectively. Each 3D volume is resized to 240 $\times$ 240 $\times$ 96 (height $\times$ width $\times$ depth). It is important to note that the datasets contain two types of image degradation functions. However, this paper focuses solely on the typical bicubic degradation \cite{Bicubic} for MRI super-resolution. Given the 2D nature of the proposed method, we generate 500 $\times$ 96 $=$ 48,000 2D training samples. (3) The ADNI dataset comprises brain MRI scans from 2,733 patients. Launched in 2003 as a public-private partnership, the initiative is led by Principal Investigator Michael W. Weiner, MD. The primary objective of ADNI has been to evaluate whether serial magnetic resonance imaging (MRI), positron emission tomography (PET), various biological markers, and clinical and neuropsychological assessments can be integrated to measure the progression of mild cognitive impairment (MCI) and early Alzheimer’s disease (AD). In this study, we utilize the TorchIO \cite{perez2021torchio} MRI artifact ecosystem, which incorporates several state-of-the-art artifact simulators designed to synthesize motion artifacts \cite{shaw2019mri}, random anisotropy \cite{billot2020partial}, and bias field inhomogeneities \cite{sudre2017longitudinal}. We randomly select 2,700 patient samples for training, with the remaining samples designated for testing and validation.

\subsection{Implementation Details}
To quantitatively analyze the model performance, we compute PSNR and SSIM \cite{SSIM} to evaluate the pixel variance between the restored and original HQ images. Following some recent works \cite{zheng2023cgc}, \cite{huang2024sparse}, \cite{mou2024empowering}, LPIPS \cite{LPIPS} and VIF \cite{VIF} are also employed to assess the perceptual reconstruction ability of each model. To evaluate the computational complexity of the proposed LRformer model, we present the parameters and multiply-accumulate operations (MACs) of all the models. As for the detailed structure of the proposed LRformer, we maintain the same depth and width as HAT \cite{HAT}. Specifically, the GFCA group number, GFCA block number, and the channel number are set to 6, 6, 180, respectively. We crop image patches sized 96 $\times$ 96 with a batch size of 4 for model training. The Adam optimizer \cite{adam} is used with $\beta_1 = 0.9$ and $\beta_2 = 0.99$ for model optimization, as well as the learning rate initialized to 2e-4 and gradually reduced to 1e-6 using CosineAnnealingRestartLR \cite{cosin} learning rate adjustment schedule. All experiments are conducted in PyTorch, utilizing NVIDIA A6000 with 48GB memory.

\begin{figure*}[!t]
\begin{minipage}[b]{1.0\linewidth}
  \centering
  \centerline{\includegraphics[width=17.8cm]{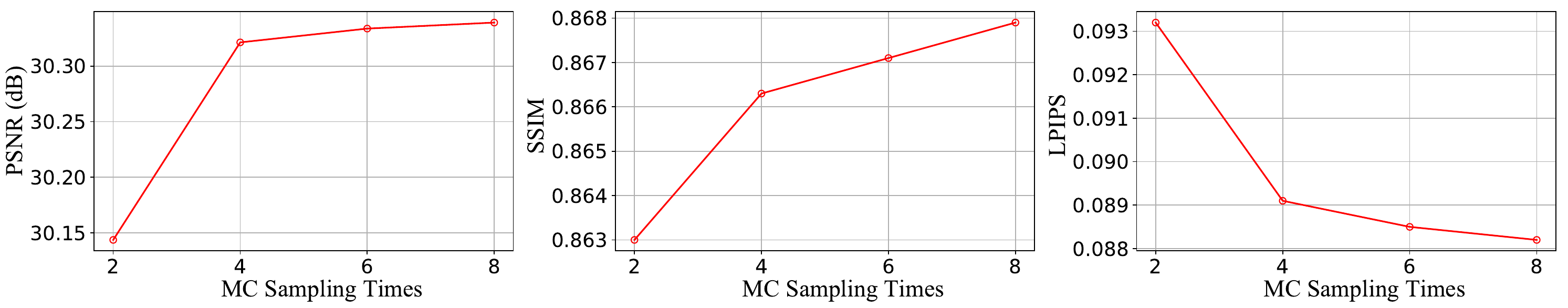}}
\end{minipage}
\caption{The image restoration performance with different Monte Carlo samples $T$. The restoration performance gains begin to plateau when $T$ reaches 6 and 8.}
\label{MCtimes}
\end{figure*}

\subsection{Comparisons with State-of-the-art Methods}
We compare our proposed LRformer model quantitatively and qualitatively with other SOTA denoising methods including MAP-NN \cite{MAPNN}, REDCNN \cite{REDCNN}, WGAN-VGG \cite{WGAN-VGG}, and SwinCT \cite{swinct}, super-resolution methods including Bicubic \cite{Bicubic}, EDSR \cite{EDSR}, RCAN \cite{RCAN}, and HAT \cite{HAT}), artifact removal methods including FBCNN \cite{FBCNN}, BlockCNN \cite{BlockCNN} and some general image restoration methods: Uformer \cite{Uformer}, ART \cite{ART}, SwinIR \cite{SwinIR} and Restormer \cite{Restormer}.

\subsubsection{Quantitative Comparisons}Table \ref{SOTA compare} presents quantitative comparisons across different MedIR tasks, including LDCT denoising, MRI super-resolution, and MRI artifact removal. 
The proposed LRformer consistently outperforms existing methods across all tasks while maintaining excellent computational efficiency. This demonstrates the effectiveness and efficiency of our approach in medical image restoration. By leveraging the lesion-semantic and shape priors of the reliable segmentation map in the frequency domain, LRformer achieves an optimal balance between computational efficiency and reconstruction performance.
Notably, the LRformer performs better not only in pixel-based metrics (PSNR, SSIM) but also in perception-based metrics (LPIPS, VIF) %in three tasks with comparable computational burden, which indicates that it is highly effective and efficient in improving the reconstruction quality of medical images by uncertainty-guided learning in the frequency domain.
across all three tasks, with a comparable computational burden. This highlights its effectiveness and efficiency in enhancing the reconstruction quality of medical images through reliable lesion-semantic prior guided learning in the frequency domain.

\subsubsection{Qualitative Comparisons}The reconstructed images from each model, visualized as in Figure \ref{visual compare}, demonstrate that the LRformer exhibits superior visual quality with clearer textures and details across all three tasks. In particular, for the LDCT denoising task, Figure \ref{visual compare} highlights that the image restored by LRformer has clearer boundaries (indicated by the yellow arrows) with reduced noise levels (indicated by the green boxes), underscoring its better noise suppression capabilities and structural preservation. %As for the MRI super-resolution task and MRI artifact removal task, it can be also observed that images recovered by the proposed ERformer are shaper and closer to the corresponding ground-truth images.
For the MRI super-resolution and MRI artifact removal tasks, the recovered images produced by LRformer are also sharper and more similar to the corresponding ground-truth images.
\subsection{Ablation Studies}
We conduct ablation studies to analyze the proposed design, which includes component analysis and evaluation of different MC sampling settings. The experiments of these ablation studies are performed using the LDCT denoising task on the AAPM dataset.

\begin{table}[!t]
\small
\begin{center}
\caption{Components ablation results of LRformer. The best results are highlighted in \textbf{bold}. w/ and w/o means ``with" and ``without", respectively.}
\label{component}
\begin{tabular*}{0.47\textwidth}{@{\extracolsep{\fill}}cccccc@{}}
\toprule
\multirow{2}*{Baseline} & \multirow{2}*{Prior}  & \multicolumn{2}{c}{ GFCA } & \multirow{2}*{PSNR ↑} & \multirow{2}*{SSIM ↑} \\ \cline{3-4} & & w/ AM & w/o AM & &  \\ \hline
        \ding{51} & \ding{55} & \ding{55} & \ding{55} & 29.4359  & 0.8502 \\
        \ding{51} & \ding{51} & \ding{55} & \ding{55} & 29.7231 & 0.8576 \\ 
        \ding{51} & \ding{51} & \ding{55} & \ding{51} & 30.1041 & 0.8633 \\ 
        \ding{51} & \ding{51} & \ding{51} & \ding{55} & \textbf{30.3213} & \textbf{0.8663} 
        \\ \bottomrule
\end{tabular*}
\end{center}
\end{table}

\subsubsection{Effectiveness of Components}
\begin{figure}[!t]
\centering
\includegraphics[width=3.4in]{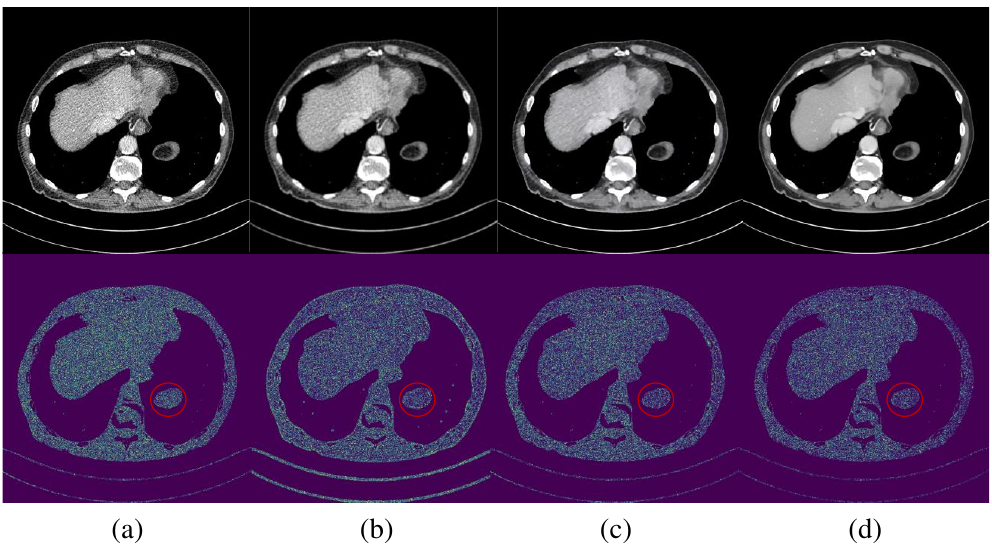}
\caption{Visualization of reconstruction results and error maps with different components. (a) baseline model. (b) with prior. (c) with prior and GFCA(w/o AM). (d) with prior and GFCA(w/ AM).}
\label{errormaps}
\end{figure}

To evaluate the effectiveness of the key components of the proposed model, we conducted four ablation experiments. For the baseline model, we replace the Guided Frequency Cross-Attention (GFCA) with vanilla vision transformer (ViT) blocks \cite{VIF} and trained the model without any priors.
Then, we integrate the reliable prior 
%by a simple matrix multiplication operation to guide the image restoration process, resulting in a 0.2872 dB PSNR improvement, which is attributed to the shape and semantic information of the handcraft prior maps. 
using matrix multiplication to guide the restoration process, resulting in a PSNR improvement of 0.2872 dB, attributable to the additional shape and lesion-semantic information from the handcrafted prior maps.
% Next, when we use GFCA (w/ AM) to integrate the prior, it further leads to a 0.3810 dB PSNR increment, which demonstrates the effectiveness of integrating priors via the frequency cross-attention mechanism. Finally, applying the adaptive mixup (AM) gains an additional PSNR bonus, which emphasizes the essential nature of contextual information interaction between the real and imaginary components. Figure \ref{errormaps} shows that the reconstruction quality is incrementally improved and the error map is gradually decreased when we gradually add these components.
Next, incorporating GFCA (w/o AM) to integrate the prior led to a further PSNR increase of 0.3810 dB, which demonstrates the benefits of using frequency domain cross-attention to incorporate prior information. Finally, applying the adaptive mixup (AM) gains an additional PSNR bonus, which emphasizes the essential nature of contextual information interaction between the real and imaginary components. Figure \ref{errormaps} illustrates the incremental improvements in reconstruction quality, gradually reducing error maps as each component is added.

\subsubsection{Setting of MC Sampling Times}
We also examine the impact of the number of MC sampling iterations $T$ on model performance, as shown in Table \ref{SamplingTimes}. 
%The results indicate a positive correlation between model performance and the number of sampling iterations $T$. Although model performance continues to improve with an increasing number of samples, a higher quantity of MC samples incurs greater computational costs. This increase in samples not only prolongs model inference time but also extends the duration required for generating consistency and discrepancy parts of the segmentation results. As the performance gains tend to saturate when $T$ is set to 6 and 8 (see supplementary material for details). Therefore, we select 4 as the optimal number of MC samplings to achieve a balance between effectiveness and efficiency.
As we can see, if the number of MC samples is too small (set as 2), it is difficult to capture the property of distribution for probabilistic embeddings, leading to unsatisfactory medical image restoration results. As we gradually increase MC sampling numbers, there is an obvious performance improvement of the proposed LRformer in terms of all metrics (PSNR, SSIM and LPIPS). Interestingly, the performance is gradually saturated when the MC samples is 6 and 8. As a result, by balancing the number of MC samples and the computational efficiency, the number of MC samples in the proposed MCD-MedSAM is set to 4.

\begin{table}[htbp]
\small
\begin{center}
\caption{The results of LRformer with different settings of MC sampling times.} 
\label{SamplingTimes}
\begin{tabular*}{0.47\textwidth}{@{\extracolsep{\fill}}cccc@{}}
\toprule
        T  & PSNR ↑ & SSIM ↑ & LPIPS ↓ \\ \hline
        2 & 30.1437 & 0.8630 & 0.0932\\
        4 & 30.3213 & 0.8663 & 0.0891 \\
        6 & 30.3337 & 0.8671 & 0.0885\\
        8 & \textbf{30.3391} & \textbf{0.8679} &\textbf{0.8882}\\
        \bottomrule
\end{tabular*}
\end{center}
\end{table}

\section{Analysis and Discussion}
\subsection{Interpretability of Frequency Learning}

\begin{figure}[!t]
\centering
% \hspace{-3.7mm}
\includegraphics[width=3.3in]{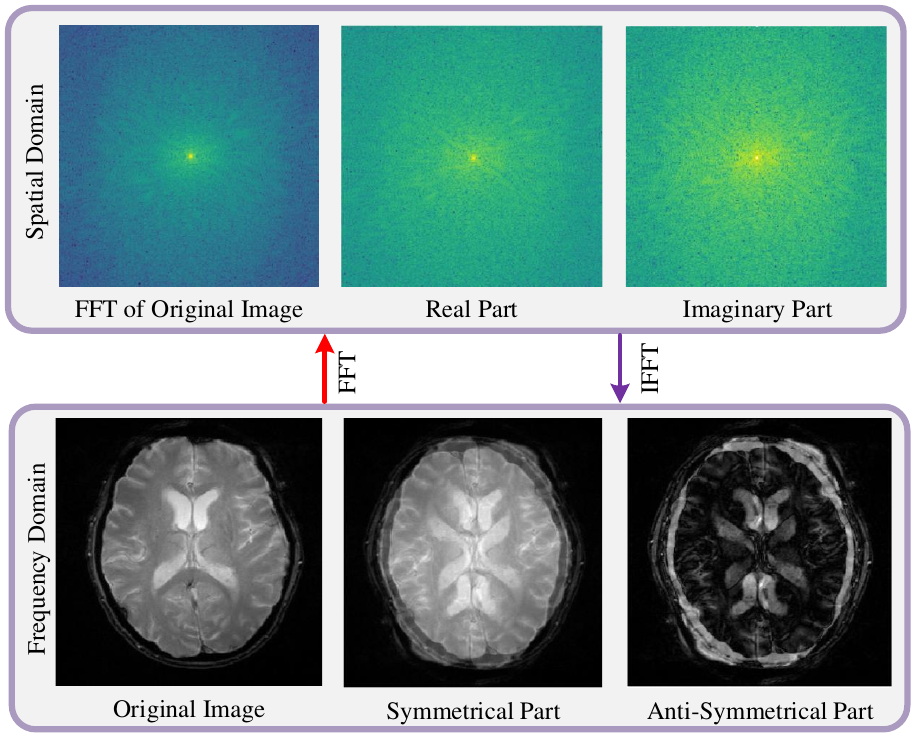}
\caption{Real symmetrical part and imaginary anti-symmetrical part of images.}
\label{frequencylearning}
\end{figure}

The conjugated symmetry of the FFT %can decompose the image into a real part and an imaginary part in the frequency domain, and these parts correspond to the symmetrical and anti-symmetrical parts in the spatial domain, respectively. 
allows the decomposition of an image into a real part and an imaginary part in the frequency domain, each corresponding to the symmetric and anti-symmetric parts in the spatial domain:
\begin{equation}
\mathcal{F}[x_{e p}(n)]=\operatorname{Re}[X(k)], \ 
\mathcal{F}[x_{o p}(n)]=j\operatorname{Im}[X(k)],
\end{equation}
where $\mathcal{F}[\cdot]$ is the FFT. $x_{e p}(n)$, $x_{o p}(n)$ denote the symmetrical and anti-symmetrical parts of the signal with sequence length $n$. $\operatorname{Re}[\cdot]$ and $\operatorname{Im}[\cdot]$ are the real and imaginary parts in the frequency domain, respectively. $j$ is the imaginary unit. Therefore, learning in the frequency domain can be interpreted as learning contextual information related to the symmetrical and anti-symmetrical parts in the spatial domain (as shown in Figure \ref{frequencylearning}). The adaptive mixup operator provides an effective way to transfer contextual information between symmetrical and anti-symmetrical components.

\subsection{Local Attribution Maps (LAMs) Analysis}
\begin{figure}[!t]
\centering
% \hspace{-3.7mm}
\includegraphics[width=3.4in]{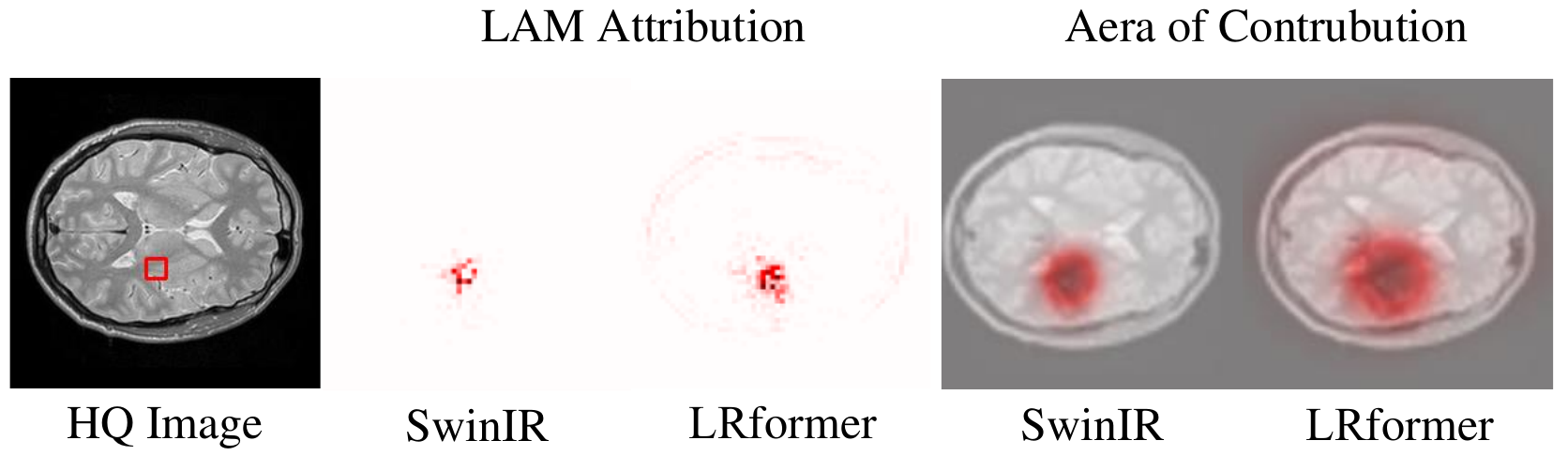}
\caption{LAM results of SwinIR and LRformer. The results indicate that SwinIR utilizes less information compared to LRformer.}
\label{LAM}
\end{figure}

As noted by Mao et al. \cite{mao2023intriguing}, the global property of FFT reveals that frequency components (for each pixel) capture the complete contextual information of spatial images. Therefore, we compare the LRformer and SwinIR using local attribution maps (LAMs) \cite{LAM}, which are used to analyze the spatial extent of utilized information for reconstruction, as shown in Figure \ref{LAM}. %Due to the global receptive field by performing contextual learning in the frequency domain, ERformer can utilize a wider range of pixels than SwinIR, which gives an interpretability perspective for model performance.
Owing to the global receptive field gained through contextual learning in the frequency domain, the LRformer can utilize a broader range of pixels compared to SwinIR, providing an interpretative explanation for its improved performance.
\section{Conclusion}

In this paper, we introduce the LRformer, a lightweight and reliable approach for MedIR tasks. By leveraging the inherent uncertainty in Bayesian Neural Networks (BNNs), we develop a reliable lesion-semantic prior producer to generate sufficiently-reliable priors. We further incorporate these priors into the restoration process by designing a guided frequency cross-attention solver, which utilizes the conjugated symmetric property of FFT to save nearly half computational complexity of the naive CA.
Extensive experiments across LDCT denoising, MRI super-resolution, and artifact removal tasks demonstrate that LRformer outperforms SOTA methods by providing clearer, more detailed reconstructions with optimal computational efficiency. Despite its remarkable performance, there remain opportunities for designing additional types of priors and enhancing the contextual correlation between the real and imaginary parts to further improve performance.


\begin{acks}
This work was supported in part by Shenzhen Science and Technology Program (Nos. \seqsplit{JCYJ20230807120010021} and \seqsplit{JCYJ20230807115959041}), National Natural Science Foundation of China (No. \seqsplit{62476052}), Sichuan Science and Technology Program (Nos. \seqsplit{2024NSFSC1473} and \seqsplit{2024ZYD0268}), and Sichuan Provincial Natural Science Foundation (No. \seqsplit{2024NSFSC0520}).
\end{acks}
\bibliographystyle{ACM-Reference-Format}
\balance
\bibliography{sample-base}


\end{document}